\documentstyle[preprint,eqsecnum,aps,epsfig]{revtex}
\tightenlines
\begin{document} 
\draft
\title{
\begin{flushright} {\small IFT-P.020/2000 \,\, gr-qc/0002072} \end{flushright}
       Low-energy sector quantization of a massless scalar field outside a 
       Reissner-Nordstrom black hole and static sources
       }

\author{ J. Casti\~neiras and G.E.A. Matsas }
\address{Instituto de F\'{\i}sica Te\'orica, 
         Universidade Estadual Paulista,\\
         Rua Pamplona 145, 01405-900, S\~ao Paulo, S\~ao Paulo, Brazil}
\def\baselinestretch{1.5}
\maketitle

\begin{abstract}

We quantize the low-energy sector of a massless scalar field in the 
Reissner-Nordstrom spacetime. This allows the analysis of processes
involving soft scalar particles occurring outside charged black holes. 
In particular, we compute the response of a static 
scalar source interacting with Hawking radiation using the Unruh 
(and the Hartle-Hawking) vacuum. This response is compared with the one 
obtained when  the source is uniformly accelerated
 in the usual vacuum of the Minkowski spacetime  
with  the same proper acceleration. We show that both responses are in general
different in opposition to the result obtained when the Reissner-Nordstrom
black hole is replaced by a Schwarzschild one.
The conceptual relevance of this result is commented.

\end{abstract}
\newpage


\section{Introduction}
\label{Intro}

We study the canonical quantization of a massless scalar
field outside a Reissner-Nordstrom black hole. This is not easy to fully 
accomplish mostly because   the explicit form of the positive and negative 
energy modes is unknown in terms of usual special functions. 
This has led many researchers to use numerical methods to analyze quantum 
field issues in this  and in similar backgrounds (see, e.g., \cite{people} 
and references 
therein).  Here we follow the  procedure developed in Ref.~\cite{HMS3'}
to  analytically quantize the low-energy sector of the scalar 
field in the Reissner-Nordstrom spacetime. This allows the analytic 
investigation of processes  involving soft  particles as, 
e.g., of the synchrotron radiation emitted by scalar sources 
orbiting charged black holes~\cite{CHM2}. 

We use our results to analyze  the following conceptual issue. 
It was recently found~\cite{HMS3} that the responses of 
(i) a static scalar source 
in the Schwarzschild spacetime with the Unruh vacuum and of (ii) 
a uniformly accelerated scalar source in the Minkowski 
spacetime with the usual vacuum are equivalent provided 
that both sources have
the same proper acceleration. It would be interesting to study, thus, whether
or not this equivalence is preserved when the Schwarzschild black hole is 
supplied with some electric charge. Because  (structureless) static
sources can only interact with {\em zero-energy} particles, we can use our 
low-energy
quantization to  answer this question accurately. Eventually we  show
that the presence of electric charge in the black hole breaks the above 
equivalence.  
This  in conjunction with the  fact that  no equivalence is found  
when the scalar field is replaced by the Maxwell one~\cite{CHM} suggests that 
the  equivalence  found in~\cite{HMS3}  is not valid, in general, 
for other spacetimes
and quantum fields. Whether or not there is
something deeper behind it, remains an open question for us. 
We will adopt natural units $\hbar = c = G = k_B = 1$ and signature 
$(+\;-\;-\;-)$.

The paper is organized as follows. In Sec.~\ref{sec:Quantization}
we quantize the low-energy sector of the massless scalar field outside a
Reissner-Nordstrom black hole. In Sec.~\ref{sec:Response} we compute
the response of a static scalar source interacting with Hawking radiation 
using the Unruh (and the Hartle-Hawking) vacuum, and compare the result 
with the one obtained when the source is uniformly accelerated in the 
Minkowski spacetime  with the usual inertial vacuum.
We present our final considerations in Sec.~\ref{sec:Discussions}. 


\section{Quantization of a massless scalar field outside a charged black hole}
\label{sec:Quantization}

The line element of a Reissner-Nordstrom black hole with mass $M$ 
and electric charge $Q\leq M$ can be written as~\cite{wald} 
\begin{equation}
     ds^2 = 
        f(r) dt^2 - f(r)^{-1} dr^2 - r^2 \left( d\theta^2 + \sin^2 
\theta d\varphi^2 \right) \;,
  \label{34} 
\end{equation}
where
\begin{equation}
     f(r) \equiv(1-r_+/r)(1-r_-/r)
  \label{34.5}
\end{equation}
and
$
r_\pm \equiv M \pm \sqrt{M^2-Q^2} \;.
$
Outside the outer event horizon, i.e. for $r > r_+$, we have a global timelike
isometry generated by the Killing field $\partial_t$. 

Let us now consider a  free massless scalar field $\Phi (x^\mu) $ 
in this background 
described by the action
\begin{equation}
     S = 
        \frac{1}{2}\int d^4x \sqrt{-g} \; 
        \nabla^\mu \Phi \nabla_\mu \Phi \;,
   \label{7}
\end{equation}
where $g \equiv {\rm det} \{ g_{\mu \nu} \}$.
In order to quantize the field, we look for a complete set 
of positive-energy solutions of the Klein-Gordon equation, 
$\Box u_{\omega l m} = 0$, in the form
\begin{equation}
     u_{\omega l m} = 
        \sqrt{\frac{\omega}{\pi}} \frac{\psi_{\omega l}(r)}{r} 
Y_{lm}(\theta,\varphi) e^{-i\omega t}  \;,
  \label{18}
\end{equation}
where $\omega\ge 0$, $l \ge 0$ and 
$m \in [-l,l]$ are the frequency
and angular momentum quantum numbers. The factor 
$\sqrt{\omega / \pi}$ was inserted for later convenience 
and $Y_{lm}(\theta,\varphi)$ are the spherical harmonics.
As a consequence $\psi_{\omega l}(r)$ must satisfy
\begin{equation}
     \left[ 
     - f(r) \frac{d}{dr} \left( f(r) \frac{d}{dr} \right) + V_{\rm eff}(r) 
     \right] \psi_{\omega l}(r) =  \omega^2 \psi_{\omega l}(r) \;,
   \label{40} 
\end{equation}
where the effective scattering potential  $V_{\rm eff}(r)$ is given by 
\begin{equation}
     V_{\rm eff}(r) = 
         \left( 
        1- \frac{2M}{r} +\frac{Q^2}{r^2} 
        \right) 
        \left( 
        \frac{2M}{r^3} -\frac{2 Q^2}{r^4}  + \frac{l(l+1)}{r^2}
        \right) \;.
   \label{41} 
\end{equation}
Note that Eq.~(\ref{40}) admits two sets of independent solutions
 which will be labeled by $\psi^\alpha_{\omega l}(r) $ with $\alpha=I, II$.
As a result, we can expand the scalar field $\Phi(x^\mu)$ in terms 
of annihilation 
$a^\alpha_{\omega l m}$and creation    
${a^{\alpha \dagger}_{\omega l m}}$ operators, as usual:
\begin{equation}
     \Phi(x^\mu) = 
        \sum_{\alpha=I,II} \sum_{l=0}^{\infty} \sum_{m=-l}^{m=+l}
        \int_0^{+\infty} d\omega 
        \left[ 
        u^\alpha_{\omega l m}(x^\mu) a^\alpha_{\omega l m} + H.c. 
        \right] \;,
   \label{10.2}
\end{equation}
where $u^\alpha_{\omega l m}(x^\mu)$ are orthonormalized according to the 
Klein-Gordon inner product~\cite{BDF}:
\begin{eqnarray}
     i\int_{\Sigma_t} d\Sigma \; n^\mu 
     \left( 
     {u^\alpha_{\omega l m}}^* \nabla_\mu u^{\alpha'}_{\omega' l' m'} - 
     \nabla_\mu {u^\alpha_{\omega l m}}^* \cdot u^{\alpha'}_{\omega' l' m'} 
     \right) 
     &=& 
     \delta_{\alpha \alpha'} \delta_{l l'} 
     \delta_{m m'} \delta(\omega - \omega') \;,
     \label{15}\\
     i\int_{\Sigma_t} d\Sigma \;  n^\mu 
     \left( 
     u^\alpha_{\omega l m} \nabla_\mu u^{\alpha'}_{\omega' l' m'} 
     - \nabla_\mu u^\alpha_{\omega l m} \cdot u^{\alpha'}_{\omega' l' m'} 
     \right) 
     &=& 0 \;.
   \label{16}
\end{eqnarray}
Here $n^\mu$ is the future-pointing unit vector normal to the volume 
element of the  Cauchy surface $\Sigma_t$. As a consequence, 
$a^\alpha_{\omega l m} $ and $a^{\alpha \dagger}_{\omega l m} $ satisfy 
simple commutation relations:
\begin{equation}
     \left[ 
     a^\alpha_{\omega l m}, {a^{\alpha' \dagger}_{\omega' l' m'}} 
     \right] = 
    \delta_{\alpha \alpha'} \delta_{l l'}
    \delta_{m m'}\delta(\omega - \omega')\;.
   \label{10.3}
\end{equation}
The Boulware vacuum $|0\rangle$ is defined by
$a^\alpha_{\omega lm}|0\rangle = 0$ for every 
$\alpha, \omega, l$ and $m$~\cite{Bo}.

\subsection{Small frequency modes}

The general solution of Eq.~(\ref{40}) in terms of special functions is 
not known. However this can be found for small frequencies as follows. 
First let us rewrite  Eq.~(\ref{40}) with $\omega = 0$ as 
\begin{equation}
     \frac{d}{dz}\left[(1-z^2)\frac{d}{dz}
     \left[ \psi_{\omega l} (y) /y \right]\right] 
     + 
    l(l+1)\left[ \psi_{\omega l} (y) /y \right] 
        = 0 \;,
   \label{45}
\end{equation}
where we have defined $y \equiv r/2M$, $y_{\pm} \equiv r_{\pm}/2M$ and
\begin{equation}
     z \equiv 
        \frac{2y - 1}{y_+ - y_-} \;.  
   \label{44}
\end{equation}
From the Legendre equation~(\ref{45}), we obtain
the  two independent solutions 
\begin{eqnarray}
    \psi_{\omega l}^I(y) & \equiv & 
       C_{\omega}^I y Q_{l}[z(y)] \;,
   \label{45.3} \\
    \psi_{\omega l}^{II}(y) & \equiv & 
       C_{\omega}^{II} y P_{l}[z(y)] \;,
   \label{45.5}
\end{eqnarray}
where $Q_{l}(z)$ and $P_{l}(z)$ are the Legendre polynomials, and
$ C_{\omega}^I$ and $ C_{\omega}^{II}$ are normalization constants.
In order to determine them, we shall analyze in more detail
the  solutions of Eq.~(\ref{40})  {\em near the horizon} and {\em at infinity},
which can be normalized for {\em arbitrary} $\omega$.

\subsection{Normal modes near the horizon and at infinity}

First let us note that by making the change of variables 
\begin{equation}
     y \to x = 
        y + \frac{ (y_+)^2 \ln\left| y-y_{+} \right| - 
        (y_-)^2\ln\left| y-y_{-}
        \right| }{y_+ - y_-} \;,
   \label{52}
\end{equation}
Eq.~(\ref{40}) takes the form
\begin{equation}
     \left[ 
     - \frac{d^2}{dx^2} + 4M^2 V_{\rm eff}[r(x)] 
     \right] \psi_{\omega l}(x) 
= 
        4M^2 \omega^2 \psi_{\omega l}(x) \;.
   \label{53}
\end{equation}
It is convenient to write the two independent solutions of Eq.~(\ref{53}) 
such that $\psi_{\omega l}^{\rightarrow}(x)$ and  
$\psi_{\omega l}^{\leftarrow}(x)$
are associated with purely incoming modes from 
the past white-hole horizon ${\cal H}^-$ 
and from the 
past null infinity ${\cal J}^-$, respectively.
These modes are orthogonal to each other with respect to the 
Klein-Gordon inner product~(\ref{15}). This can be  seen
by choosing $\Sigma_t ={\cal H}^- \cup {\cal J}^- $ 
in Eq.~(\ref{15})  and recalling that 
$\psi_{\omega l}^{\rightarrow}(x)$ 
and 
$\psi_{\omega l}^{\leftarrow}(x) $ 
vanish on $ {\cal J}^-$ and ${\cal H}^-$, respectively.
Hence, by noting  from Eq.~(\ref{41}) that  
close  ($x<0, |x| \gg 1 $) to and far away ($x \gg 1 $) from the horizon, 
the scattering potential becomes 
$V_{\rm eff}(r) \approx 0$ and  $V_{\rm eff}(r) \approx l(l+1)/r^2$, 
respectively, we write 
\begin{equation}
     \psi_{\omega l}^{\rightarrow}(x) \approx 
        \left\{ 
          \begin{array}{lc}
         A_{\omega l} \left( e^{2iM\omega x} + 
         {\cal R}_{\omega l}^{\rightarrow} 
e^{-2iM\omega x} \right) & (x < 0\;, |x| \gg 1) , \\
          2 i^{l+1} A_{\omega l}   
          {\cal T}_{\omega l}^{\rightarrow} M\omega x 
h_l^{(1)}(2M\omega x) & (x \gg 1) ,
          \end{array} 
        \right.   
   \label{51}
\end{equation}
and
\begin{equation}
     \psi_{\omega l}^{\leftarrow}(x) \approx 
        \left\{ 
          \begin{array}{lc}
            B_{\omega l} {\cal T}_{\omega l}^{\leftarrow} e^{-2iM\omega x} 
            & (x< 0, |x| \gg 1), \\
            B_{\omega l} \left[ 2(-i)^{l+1} M\omega x 
            {h_l^{(1)}(2M\omega x)}^* 
  + 2 i^{l+1} {\cal R}_{\omega l}^{\leftarrow} M\omega x h_l^{(1)} 
  (2M\omega x) 
            \right] & (x \gg 1).
          \end{array} \right. 
   \label{68}
\end{equation}
Here $h_l^{(1)}(2M \omega x)$  
are the spherical Hankel functions and
$
\left| {\cal R}_{\omega l}^{\leftarrow} \right|^2 ,
\left| {\cal R}_{\omega l}^{\rightarrow} \right|^2 
$  
and
$
\left| {\cal T}_{\omega l}^{\leftarrow} \right|^2 ,
\left| {\cal T}_{\omega l}^{\rightarrow} \right|^2
$
are the reflection and transmission coefficients, respectively, 
satisfying the usual probability conservation equations: 
$
\left| {\cal R}_{\omega l}^\rightarrow \right|^2 
+ 
\left| {\cal T}_{\omega l}^\rightarrow \right|^2 
= 1 
$
and 
$
\left| {\cal R}_{\omega l}^\leftarrow \right|^2 
+ 
\left| {\cal T}_{\omega l}^\leftarrow \right|^2
= 1 
$.
Note that $h_l^{(1)} \approx (-i)^{l+1} \exp(ix)/x $ for $|x| \gg 1$.  
The  normalization constants $A_{\omega l}$ and $B_{\omega l}$
are obtained (up to an arbitrary phase) by letting
normal modes~(\ref{18}) in the Klein-Gordon inner product~(\ref{15})
and using Eq.~(\ref{53}) to transform the integral into a surface
term:
\begin{equation}
          \left. \frac{1}{\omega - \omega'} 
          \left[\psi_{\omega l}(x) \frac{d}{dx} 
\psi_{\omega' l}^{*}(x) - \psi_{\omega' l}^{*}(x) \frac{d}{dx} \psi_{\omega 
l}(x) \right] \right|^{x \to +\infty}_{x \to -\infty} 
= 
\frac{2\pi M}{\omega} \delta (\omega-\omega') \;.
   \label{57}
\end{equation}
By using the asymptotic solutions~(\ref{51})-(\ref{68}) in Eq.~(\ref{57}),
we obtain $A_{\omega l} = B_{\omega l}= (2\omega)^{-1}$.

\subsection{Normalization constants}

Now we are able to determine the normalization constants 
$C_{\omega}^I$ and $C_{\omega}^{II}$ 
by comparing Eqs.~(\ref{45.3})-(\ref{45.5}) close and 
far away from the black hole
with our normalized functions~(\ref{51})-(\ref{68}) 
in the low-frequency regime
($2M\omega x \ll 1$).  Let us begin noticing that for $2M\omega x \ll 1$,
we have near the horizon [see Eq.~(\ref{51})]
\begin{equation}
     \psi_{\omega l}^{\rightarrow}(x) \approx 
        M x \left[ 
                   \frac{(1+{\cal R}_{\omega l}^{\rightarrow})}{2M\omega x}
                  +i(1-{\cal R}_{\omega l}^{\rightarrow})
             \right]     
         \;\;\;\;\; (x < 0\;, |x| \gg 1) .
   \label{61.5}
\end{equation}
In order that Eq.~(\ref{61.5}) has a good behavior in the low-frequency regime 
we conclude that ${\cal R}_{\omega l}^{\rightarrow} \approx -1 + 
{\cal O} (\omega)$. 
As a consequence, for  $2M\omega x \ll 1$ we obtain from Eq.~(\ref{61.5}) 
that  
\begin{equation}
     \psi_{\omega l}^{\rightarrow}(x) \approx
       2iMx       \;\;\;\;\; (x < 0\;, |x| \gg 1) \;.
   \label{65}
\end{equation}
Now, we recall that in the low-frequency 
regime $\psi_{\omega l}^{\rightarrow}(x)$ 
is mostly reflected by the 
scattering potential back 
to the horizon  and thus cannot 
be associated  with $\psi_{\omega l}^{II}(x)$ 
which {\em grows} asymptotically [see Eq.~(\ref{45.5}) and recall that
$P_l(z) \sim z^l$ as $z\gg 1 $ ($r \gg r_+$)]. This is not so 
for $\psi_{\omega l}^{I}(x)$ which decreases asymptotically and 
indeed fits $\psi_{\omega l}^{\rightarrow}(x)$.  
This can be shown as follows. Let us first note that for $z \approx 1$ 
($r \approx r_+$)
\begin{eqnarray}
     Q_{l}(z) & \approx &
        \frac{1}{2} \ln\left| \frac{z+1}{z-1} \right| - \sum_{k=1}^{l} 
\frac{1}{k} \nonumber \\
              & \approx &
         \frac{\left[-x + y_+ +\ln(y_+ - y_-) \right] (y_+ - y_-)}{2{y_+}^2} - 
\sum_{k=1}^{l} \frac{1}{k} 
   \label{62}
\end{eqnarray}
where we have used Eqs.~(\ref{44}) and (\ref{52}).
Thus, close to the horizon, we obtain from  Eq.~(\ref{45.3}) that
\begin{equation}
     \psi_{\omega  l}^I (x) \approx
       - C_{\omega}^I \frac{(y_+ - y_-)}{2y_{+}}x 
         \;\;\;\;\; (x < 0\;, |x| \gg 1) \;.
   \label{64}
\end{equation}
Comparing Eqs.~(\ref{64}) and (\ref{65}) we find the normalization constant
\begin{equation}
      C_{\omega}^I = -4iM y_{+}/(y_+ - y_-) \;.
   \label{66}
\end{equation}
Therefore, we write from Eq.~(\ref{45.3}) 
\begin{equation}
     \psi_{\omega l}^I (x) =\frac{ -4iM y_{+} y Q_{l}[z(y)]}{y_+ - y_-}  \;,
   \label{67}
\end{equation}
and from Eq.~(\ref{18}), we obtain the corresponding normalized 
low-frequency modes (up to an arbitrary phase):
\begin{equation}
     u_{\omega lm}^I (x^\mu)=
                \frac{ 2  y_{+}\; \omega^{1/2}}{\pi^{1/2}\; (y_+ - y_-) } 
               Q_{l}[z(x)] Y_{lm}(\theta,\varphi) e^{-i\omega t} \;.
   \label{uI}
\end{equation}
Now we fit $\psi_{\omega l}^{I}(x)$ and $\psi_{\omega l}^{\rightarrow}(x)$
asymptotically to determine the low-frequency 
transmission coefficient $|{\cal T}_{\omega l}^{\rightarrow}|^2$ 
[see Eq.~(\ref{51})].
For $x\gg 1$, Eq.~(\ref{67}) becomes
\begin{equation}
     \psi_{\omega l}^I (x) \approx
     \frac{-2iM(l!)^2 y_+ (y_+ -y_-)^{l} x^{-l}}{(2l+1)!}
     \;\;\;\;\; (2M\omega x \ll 1) \;,
\label{67.1}
\end{equation}
where we have used that in this region
$$
Q_l [2y/(y_+ -y_-)] \approx 
\frac{(l!)^2 (y_+ -y_-)^{l+1} y^{-l-1}}{2 (2l+1)! }\;.
$$
Now, from Eq.~(\ref{51}), we have in the low-frequency regime
and for $x \gg 1$ that
\begin{equation}
     \psi_{\omega l}^\rightarrow (x) \approx 
     \frac{i^{l}  (2l)!\; {\cal T}^\rightarrow_{\omega l} x^{-l}}
          {2^{2l+1} l! M^l \omega^{l+1}}
     \;\;\;\;\; (2M\omega x\ll 1) \;,
\label{psi->}
\end{equation}
where we have used that 
\begin{equation}
h_l^{(1)}(2M \omega x)  =  j_l(2M \omega x) + i n_l(2M \omega x)
\label{h1}
\end{equation} 
and the fact that the spherical Bessel and Newman functions satisfy 
 (see Eq.~(11.156) of Ref.~\cite{arfken})
\begin{equation}
     j_l(2M \omega x)\approx  \frac{2^l l!}{(2l+1)!} (2M \omega x)^l  
\label{69.25}
\end{equation}
and
\begin{equation}
     n_l(2M \omega x) \approx - \frac{(2l)!}{2^l l!} 
     (2M \omega x)^{-(l+1)} \;,
   \label{69.3}
\end{equation}
respectively, for $2M \omega x \ll 1$.
Thus Eqs.~(\ref{67.1}) and (\ref{psi->}) coincide provided that
\begin{equation}
{\cal T}^\rightarrow_{\omega l} = 
 \frac{ 2^{2l+2} (-i)^{l+1} y_+ (y_+ -y_-)^l (l!)^3 (M\omega)^{l+1}}
      { (2l+1)! (2l)!} \;.
\label{tau->}
\end{equation}
(Eventually this will be also used as a consistency 
check for our calculations.)

Now, let us turn our attention to $\psi_{\omega l}^{\leftarrow}(x)$ 
which should be fitted with $\psi_{\omega l}^{II}(x)$. Note that 
$\psi_{\omega l}^{I}(x)$ grows close to the horizon and so
cannot be associated with low-frequency left-moving modes which must be 
mostly reflected back to infinity by the scattering potential
(see Eq.~(\ref{45.3}) and recall that $Q_l(z) \approx -\log|z-1|^{1/2}$
as $z\approx 1$).
In order to fit $\psi_{\omega l}^{\leftarrow}(x)$ and
$\psi_{\omega l}^{II}(x)$   asymptotically, we must 
use Eqs.~(\ref{h1}) and (\ref{69.25})-(\ref{69.3}) in 
Eq.~(\ref{68}) for $x \gg 1$.
Moreover it turns out that  this compatibility
is achieved if and only if 
${\cal R}^\leftarrow_{\omega l} \approx (-1)^{l+1}$. 
As a result we obtain
\begin{equation}
     \psi_{\omega l}^{\leftarrow}(x) 
      \approx  
          \frac{2^{2l+1}(-i)^{l+1}  l!\; \omega^l (Mx)^{l+1}}{(2l+1)!}    
     \;\;\;\;\; (x \gg 1) 
   \label{77.5}
\end{equation}
for $2M\omega x \ll 1$.
Now,  we  note that $P_l(z)\approx \left[(2l)!/2^l (l!)^2 \right] z^l $ for 
$z \gg 
1$ (see Eqs. (8.837.2) and (8.339.2) of Ref.~\cite{grad}). Hence, using 
Eqs.~(\ref{44}) and (\ref{45.5}), we find that
\begin{equation}
     \psi_{\omega l}^{II} (x) \approx  C_{\omega}^{II}  
     \frac{(2l)! y^{l+1}}{(l!)^2 (y_+ - y_-)^l} \;\;\;\;\;\; (x \gg 1) \;.
   \label{71}
\end{equation}
Comparing this equation with Eq.~(\ref{77.5}) and recalling that 
$x \approx y$
at infinity, we find the normalization constant
\begin{equation}
       C_{\omega}^{II} = \frac{2^{2l+1}(-i)^{l+1}(l!)^3 M^{l+1} (y_+ - y_-)^l 
\omega^l}{(2l+1)!(2l)!} \;.
   \label{78}
\end{equation}
Therefore 
\begin{equation}
     \psi_{\omega l}^{II}(x) = 
     \frac{2^{2l+1}(-i)^{l+1}(l!)^3 M^{l+1} 
     (y_+ - y_-)^l  \omega^l y P_{l}[z(y)]}{(2l+1)!(2l)!}  
   \label{78.5}
\end{equation}
and the corresponding normalized small frequency modes are (up to an 
arbitrary phase)
\begin{equation}
     u_{\omega lm}^{II}(x^\mu) = \frac{2^{2l}(l!)^3 M^{l} (y_+ - y_-)^l 
\omega^{l+1/2}}{\pi^{1/2} (2l+1)!(2l)!}  P_{l}[z(x)] Y_{lm}(\theta,\varphi) 
e^{-i\omega t} \;.
   \label{uII}
\end{equation}
It can be directly verified that by fitting Eq.~(\ref{78.5}) 
close to the horizon 
with Eq.~(\ref{68}) for $ 2M\omega x \ll 1$, we
obtain ${\cal T}^\leftarrow_{\omega l} = {\cal T}^\rightarrow_{\omega l}$ 
[see Eq.~(\ref{tau->})], as indeed required for consistency.  
Clearly this guaranties that 
$ |{\cal R}^\leftarrow_{\omega l}| = |{\cal R}^\rightarrow_{\omega l}| $.
Note, however, that  ${\cal R}^\leftarrow_{\omega l}$ and
${\cal R}^\rightarrow_{\omega l}$
will in general differ by a phase  (in contrast to
${\cal T}^\leftarrow_{\omega l}$ and ${\cal T}^\rightarrow_{\omega l}$).

Eq.~(\ref{10.2}) in conjunction with 
Eqs.~(\ref{uI}) and (\ref{uII}) conclude our low-frequency sector 
quantization.


\section{Response of a static scalar source interacting with Hawking radiation}
\label{sec:Response}

Let us now  compute the response of a static  
source  to the Hawking radiation in the Reissner-Nordstrom spacetime. 
We will consider both  Unruh and Hartle-Hawking vacua.
Let us describe our pointlike  scalar source lying at $(r_0, \theta_0, 
\varphi_0)$ by
\begin{equation}
     j(x^\mu) = \frac{q}{\sqrt{-h\;}} \delta(r-r_0) \delta(\theta-\theta_0) 
     \delta(\varphi-\varphi_0) \;,
     \label{80}
\end{equation}
where  $q$ is a small coupling constant and 
$h =- f^{-1} r^4 \sin^2 \theta$ 
is the determinant of the spatial metric 
induced over the equal time hypersurface $\Sigma_t$. Note that Eq~(\ref{80})
guaranties that
\begin{equation}
     \int_{\Sigma_t} d\Sigma \; j = q 
   \label{81}
\end{equation}
wherever the source lyies. Let us now couple our source $j(x^\mu)$ to 
a massless scalar field $\Phi (x^\mu) $ as described by the interaction action
\begin{equation}
     S_I = 
       \int d^4x \sqrt{-g}\; j\; \Phi \;.
  \label{A7}
\end{equation}

The total source response, i.e., total particle emission and
absorption probabilities per proper time associated with the source,
is given by
\begin{equation}
     R \equiv  
       \sum_{\alpha=I,II} 
       \sum_{l=0}^{\infty} 
       \sum_{m=-l}^{l} 
       \int_0^{+\infty} d\omega  R^\alpha_{\omega l m}  \;  ,
   \label{r35}
\end{equation}
where 
\begin{equation}
     R^\alpha_{\omega lm} \equiv 
     \tau^{-1}
     \left\{ 
 \left|{{\cal A}^\alpha_{\omega lm}}^{\rm em}  \right|^2 [1+ n^\alpha(\omega)] 
+ \left|{{\cal A}^\alpha_{\omega lm}}^{\rm abs} \right|^2 n^\alpha(\omega) 
     \right\}
   \label{82}
\end{equation}
and $\tau$ is the  source's total proper time. (This is well defined
since our source  is pointlike.) Here
$
{{\cal A}^\alpha_{\omega lm}}^{\rm em} 
\equiv 
\left\langle \alpha \omega l m \left| S_I \right| 0 \right\rangle
$
and
$
{{\cal A}^\alpha_{\omega lm}}^{\rm abs} 
\equiv 
\left\langle 0 \left| S_I \right| \alpha \omega l m \right\rangle
$
are the emission and absorption amplitudes,   respectively,
of Boulware states  $|\alpha \omega lm\rangle$, at the tree level. 
Moreover
\begin{equation}
n_U^\alpha(\omega)
\equiv 
\left\{ 
        \begin{array}{cl}
      ( e^{\omega\beta } - 1 )^{-1} & \;\;\; {\rm for} \;\;\; \alpha = I \;,\\
      0                           & \;\;\; {\rm for} \;\;\; \alpha = II \;,
          \end{array} 
\right.
\label{nU}
\end{equation}
and     
\begin{equation}
n_{HH}^\alpha(\omega)
\equiv 
\left\{ 
          \begin{array}{cl}
       (e^{\omega\beta } - 1 )^{-1} & \;\;\; {\rm for} \;\;\; \alpha = I \;,\\
       (e^{\omega\beta } - 1 )^{-1} & \;\;\; {\rm for} \;\;\; \alpha = II \;,
          \end{array} 
\right.
\label{nHH}
\end{equation}
for the Unruh and Hartle-Hawking vacua, respectively, with
\begin{equation}
          \beta^{-1} = \frac{y_+-y_-}{8\pi My_+^2}
      \;.
   \label{39}
\end{equation}
We recall that the Unruh vacuum is characterized by a thermal flux leaving  
${\cal H}^-$
with Hawking temperature $\beta^{-1}$ at  infinity given by Eq.~(\ref{39}) 
while the Hartle-Hawking vacuum has in addition   
a thermal flux coming from  ${\cal J}^-$ characterized by the same
temperature at infinity~\cite{HaU}.
 
Let us note that  because structureless static sources~(\ref{80}) 
can only interact with 
{\em zero-energy} modes,  the total response 
of this  source  
in the Boulware vacuum  vanishes. This is not so, 
however, in the presence of a background thermal bath since the 
absorption  and (stimulated) emission rates render it non-zero.
In order to deal with zero-energy modes, we need a ``regulator'' to avoid the
appearance of intermediate indefinite results. (For a more comprehensive
discussion on the interaction of static sources with zero-energy modes, 
see Ref.~\cite{HMS}.)
For this purpose we let the  coupling constant $q$ to smoothly oscillate 
with frequency $\omega_0$, writing Eq.~(\ref{80}) in the form
\begin{equation}
       j_{\omega_0}(x^\mu) = 
         \frac{ q_{\omega_0}}{\sqrt{-h\;}} \delta(r-r_0) 
\delta(\theta-\theta_0) \delta(\varphi-\varphi_0) \;,
\label{A8}
\end{equation}
where $ q_{\omega_0} \equiv \sqrt{2} q \cos(\omega_0 t)$ 
and  taking the limit 
$\omega_0 \rightarrow 0$ at the end. The factor $\sqrt{2}$ 
has been introduced to guaranty that the time average
$\left\langle  |q_{\omega_0}(t)|^2\right\rangle_t = q^2$ 
since at the tree level the absorption and emission rates 
are functions of $q^2$.
By using Eqs.~(\ref{A8}) and~(\ref{10.2}) in~(\ref{A7}) 
we obtain the following
absorption amplitude 
\begin{equation}
{{\cal A}^\alpha_{\omega lm}}^{\rm abs} = 
q\sqrt{2\pi \omega_0} (\psi^\alpha_{\omega_0 l} (r_0)/r_0) f^{1/2} (r_0)
Y_{lm} (\theta_0,\varphi_0)
 \delta(\omega - \omega_0)  \;,
\label{amplitude}
\end{equation}
and we recall that 
$|{{\cal A}^\alpha_{\omega lm}}^{\rm em}| = 
|{{\cal A}^\alpha_{\omega lm}}^{\rm abs}| $.
By letting Eq.~(\ref{amplitude})  into  Eq.~(\ref{82}) we obtain 
\begin{equation}
{R^\alpha_{\omega lm}} = 
q^2 \omega_0 (|\psi^\alpha_{\omega_0 l}(r_0)|^2/r_0^2) f^{1/2} (r_0)
|Y_{lm} (\theta_0,\varphi_0)|^2 (1+2n^\alpha(\omega_0))
 \delta(\omega - \omega_0) \;,
\label{response}
\end{equation}
where it was used that the source's total proper time is 
$\tau = 2\pi f^{1/2} (r_0) \lim_{\omega \to 0} \delta (\omega)$.
(Here $f^{1/2} (r_0)$ is the gravitational red-shift factor.)

Let us first consider the Unruh vacuum. By using
Eqs.~(\ref{67}), (\ref{nU}) and~(\ref{response})  
in Eq.~(\ref{r35}) and making $\omega_0 \to 0$
at the end, we compute the total response 
\begin{equation}
     R_{U} = \frac{q^2 a {(M-Q^2/{r_{+}})}}{4\pi^2 {(M-Q^2/{r_{0}})}}  
   \label{r35.5}
\end{equation}
(note that modes $u^{II}_{\omega lm} (x)$ do not give any contribution
here), where 
$$
a = \frac{f^{-1/2} (r_0)}{2}\frac{df(r_0)}{dr_0}  
$$
is the  source's proper acceleration and we have used
\begin{equation}
     \sum_{m=-l}^{l} \left|Y_{lm}(\theta_0,\varphi_0)\right|^2 = 
\frac{2l+1}{4\pi}
   \label{r33}
\end{equation}
and~\cite{HMS3}
\begin{equation}
     \sum_{l=0}^{\infty} \left| Q_{l}(s)\right|^2 (2l+1) = 
     \frac{1}{s^2-1} \;.
   \label{r34}
\end{equation}
 Next we compare Eq.~(\ref{r35.5})  with 
\begin{equation}
     R_{M} = \frac{q^2 a}{4\pi^2}  \;, 
   \label{0}
\end{equation}
which is the response associated with our scalar source when it is uniformly
accelerated in the usual vacuum of the Minkowski spacetime
with proper acceleration $a$. 
We note that although Eqs.~(\ref{r35.5}) and~(\ref{0}) coincide when $Q=0$,
as found in Ref.~\cite{HMS3}, they do not for $Q\neq 0$.
As a result, the presence of electric charge inside the black hole
breaks the response equivalence. 

We note that  the
equality between  Eqs.~(\ref{r35.5}) and~(\ref{0}) is  recovered when
$r_0 \approx r_+$. Hence
close to the horizon, a static source in the Unruh vacuum responds
as if it were static in the Rindler wedge 
(i.e., uniformly accelerated in the Minkowski
spacetime) with the usual inertial vacuum
provided that both sources have the same proper acceleration. 
Moreover, Eq.~(\ref{r35.5}) can be written  in this region in terms
of the  proper temperature~\cite{To} 
$\beta_0^{-1}= \beta^{-1}/ \sqrt{f(r_0)}$ on the source's location as
\begin{equation}
     R_U \approx \frac{q^2}{2\pi \beta_0} \;.
  \label{a0}
\end{equation} 
Eq.~(\ref{a0}) coincides with the response associated with our source
when it is at rest  in the Minkowski spacetime 
with a background thermal bath 
characterized by a temperature $\beta^{-1}_0$. 
This result is not surprising because
close to the horizon the scattering potential 
vanishes and the zero-energy
modes leaving ${\cal H^-}$ are completely  
reflected back towards the horizon. 

Now let us turn our attention to the Hartle-Hawking vacuum.
An  analogous calculation leads us to the following response:
\begin{equation}
     R_{HH} = \frac{q^2 a\; {(M-Q^2/{r_{+}})}}{4\pi^2{(M-Q^2/{r_{0}})}}  
               + \frac{q^2 (M - Q^2/r_+)( M - Q^2/r_0)}{4 \pi^2 r_+^2 r_0^2\; 
               a } \;,
   \label{a4}
\end{equation}
where we have used that $P_0[z(r_0)]= 1$ and $Y_{00}  =  {1}/\sqrt{4\pi} $.
[Note that, in this case, only $l=0$ contributes in Eq.~(\ref{r35}).]
The first term in the right-hand side of Eq.~(\ref{a4}) is identical to the 
one obtained with the Unruh vacuum
and is associated with the thermal flux leaving ${\cal H}^-$.
The second term is associated with the thermal flux coming from 
${\cal J}^-$. As a consistency check  we note that for $r \to r_+$, we obtain 
$R_{HH}=R_U$. This should be so because close to the horizon, zero-energy
particles  coming from  ${\cal J}^-$ cannot
overpass the scattering barrier. Consequently, in this limit,
the second term in the right-hand 
side of Eq.~(\ref{a4}) must vanish. Now, when
the source is far away from the hole, only the second term 
in the right-hand side of  Eq.~(\ref{a4}) contributes because zero-energy
particles leaving ${\cal J}^-$ are not able to reach the asymptotic region.
Moreover, in this region,  Eq.~(\ref{a4}) can be rewritten in the form
\begin{equation}
     R_{HH} \approx \frac{q^2}{2\pi \beta} \;.
  \label{a0'}
\end{equation} 
Hence, far away from the hole, the source behaves as if it were 
in the Minkowski spacetime immersed
in a thermal bath  with temperature  $\beta^{-1}$, as expected.


\section{Discussions}
\label{sec:Discussions}

We have quantized the low-energy sector of  a massless scalar field  
in the Reissner-Nordstrom spacetime.  
The results obtained were used to analyze  the response of a static
source interacting with  Hawking radiation using the 
Unruh and the Hartle-Hawking vacua.
We have shown  that, in general, static sources outside 
{\em charged} black holes (with 
the Unruh vacuum) do not behave  similarly to uniformly accelerated sources 
in the Minkowski spacetime (with the usual inertial vacuum) 
as previously found for
{\em neutral} black holes~\cite{HMS3}. 
This  in conjunction with the  fact that  no equivalence is found  
when the scalar field is replaced by the Maxwell one~\cite{CHM} 
shows that 
the  equivalence  found in~\cite{HMS3}  is not valid, in general, 
for other spacetimes
and quantum fields.
 Whether or not there is something deeper behind it, remains an open 
question for us.  We have also verified that close to  and  
far away from the horizon
our source behaves as if it were at rest in a thermal bath  in the
Minkowski spacetime with proper temperature associated with the   
Unruh and Hartle-Hawking vacua, respectively. The low-energy quantization
presented here can be used to analyze  other processes 
occurring outside charged black hole.

\acknowledgments
J.C. and G.M. would like to acknowledge full and partial support from the 
Funda\c{c}\~ao de Amparo \`a Pesquisa do Estado de S\~ao Paulo (FAPESP) 
and Conselho Nacional de Desenvolvimento Cient\'{\i}fico  e Tecnol\'ogico 
(CNPq), respectively.

\end{document}